\documentclass[journal]{IEEEtran}
\makeatletter
\def\ps@headings{%
\def\@oddhead{\mbox{}\scriptsize\rightmark \hfil \thepage}%
\def\@evenhead{\scriptsize\thepage \hfil \leftmark\mbox{}}%
\def\@oddfoot{}%
\def\@evenfoot{}}
\makeatother
\IEEEoverridecommandlockouts

\usepackage{booktabs} 
\usepackage{graphicx}
\usepackage{caption}
\usepackage{subcaption}
\usepackage{url}
\usepackage{epstopdf}
\usepackage{amsmath}
\usepackage{amssymb}
\usepackage{amsthm}
\usepackage{amsfonts}
\usepackage{float}
\usepackage{rotating}
\usepackage{array}
\usepackage{bm}
\usepackage{multirow}
\usepackage{tabu}
\usepackage[n,advantage,operators,sets,adversary,landau,probability,notions,logic,ff,mm,primitives,events,complexity,asymptotics,keys]{cryptocode}
\usepackage{thmtools, thm-restate}
\usepackage{booktabs} 
\usepackage{tikz}

\usepackage{cite}
\usepackage{textcomp}
\usepackage{xcolor}

\def\BibTeX{{\rm B\kern-.05em{\sc i\kern-.025em b}\kern-.08em
    T\kern-.1667em\lower.7ex\hbox{E}\kern-.125emX}}

\newtheorem{theorem}{Theorem}
\newtheorem{assumption}{Assumption}
\theoremstyle{definition}
\newtheorem{definition}{Definition}

\DeclareMathAlphabet{\mathbfit}{OML}{cmm}{b}{it}
\DeclareMathAlphabet{\mathbfsf}{OML}{cmm}{b}{s}

\newcommand{\para}[1]{{\smallskip\noindent\textbf{#1~}}}

\newcommand{\savesp}{\vspace{-2mm}}

\begin{document}

\title{Cumulative Message Authentication Codes for Resource-Constrained IoT Networks}

\author{He Li, Vireshwar Kumar,~\IEEEmembership{Member,~IEEE,} Jung-Min (Jerry) Park,~\IEEEmembership{Fellow,~IEEE,} and Yaling Yang,~\IEEEmembership{Member,~IEEE}
\thanks{A preliminary version of some portions of this work appeared in \cite{Li20}.}
\thanks{H. Li, J. Park, and Y. Yang are with the Department of Electrical and Computer Engineering, Virginia Tech, Blacksburg, Virginia, USA (e-mail: heli@vt.edu, jungmin@vt.edu, yyang8@vt.edu).}
\thanks{V. Kumar is with the Department of Computer Science and Engineering, Indian Institute of Technology, Delhi, India (e-mail: viresh@cse.iitd.ac.in).}
}

\maketitle

\begin{abstract}
In resource-constrained IoT networks, the use of conventional message authentication codes (MACs) to provide message authentication and integrity is not possible due to the large size of the MAC output. A straightforward yet naive solution to this problem is to employ a truncated MAC which undesirably sacrifices cryptographic strength in exchange for reduced communication overhead. In this paper, we address this problem by proposing a novel approach for message authentication called \textit{Cumulative Message Authentication Code} (CuMAC), which consists of two distinctive procedures: \textit{aggregation} and \textit{accumulation}. In aggregation, a sender generates compact authentication tags from segments of multiple MACs by using a systematic encoding procedure. In accumulation, a receiver accumulates the cryptographic strength of the underlying MAC by collecting and verifying the authentication tags. Embodied with these two procedures, CuMAC enables the receiver to achieve an advantageous trade-off between the cryptographic strength and the latency in processing of the authentication tags. Furthermore, for some latency-sensitive messages where this trade-off may be unacceptable, we propose a variant of CuMAC that we refer to as \textit{CuMAC with Speculation} (CuMAC/S). In addition to the aggregation and accumulation procedures, CuMAC/S enables the sender and receiver to employ a speculation procedure for predicting future message values and pre-computing the corresponding MAC segments. For the messages which can be reliably speculated, CuMAC/S significantly reduces the MAC verification latency without compromising the cryptographic strength. We have carried out comprehensive evaluation of CuMAC and CuMAC/S through simulation and a prototype implementation on a real car.

\end{abstract}


\section{Introduction}
\label{sec:introduction}
In emerging applications, such as intelligent automobiles, industrial control systems and smart city networks, a large number of \emph{energy-constrained} computing devices are getting closely integrated with the existing computer infrastructure through \emph{bandwidth-constrained} networks to form the Internet-of-Things (IoT) \cite{Raz17}. The successful adoption of those applications will partially depend on our ability to thwart security and privacy threats, including message forgery and tampering. Today, message authentication code (MAC) is the most commonly used method for providing message authenticity and integrity in wired/wireless network applications. To employ MACs in a resource-constrained (i.e., energy and/or bandwidth-constrained) network, we need to consider two problems: the computational burden on the devices for generating/verifying the MAC, and the additional communication overhead incurred due to the inclusion of the MAC in each message frame/packet. The first problem can be addressed by using dedicated hardware and cryptographic accelerators \cite{Esc09,Soj14}. However, the second problem is not as easy to address.

\para{Problem.}
The cryptographic strength of a MAC depends on the cryptographic strength of the underlying cryptographic primitive (e.g. a hash or block cipher), the size and quality of the key, and the size of the MAC output. Hence, a \textit{conventional MAC} scheme typically employs at least a few hundred bits of MAC output to ensure a sufficient level of cryptographic strength. Unfortunately, in resource-constrained IoT networks (e.g., energy-constrained low-power wide-area network with battery-powered devices and bandwidth-constrained in-vehicle controller area network), the payload size of each packet is very short, i.e., less than a hundred bits~\cite{Nil08}. As such, not more than a few bits can be spared to include an \textit{authentication tag}, prohibiting the usage of the conventional MAC~\cite{Raz17}.

\para{Related Work.}
The legacy solution for generating a short authentication tag is to truncate the output of a conventional MAC so that it fits a message packet~\cite{Wan17,Szi08, sch11}. This type of MAC is called a \textit{truncated MAC}. However, the truncated MAC sacrifices cryptographic strength in exchange for reduced communication overhead and energy consumption, which may be undesirable, or even unacceptable, in some applications. Note that the truncated MAC without sufficient cryptographic strength renders the application vulnerable to collision attacks \cite{bhargavan2016transcript}. To enable authentication with enhanced cryptographic strength, Katz et al. propose the concept of \emph{aggregate MAC} where conventional MACs of multiple messages are combined into one aggregate MAC, and transmitted over successive packets \cite{Kat08}. Similarly, Nilson et al. propose a \emph{compound MAC} which is calculated on a compound of multiple messages, and distributed over successive packets \cite{Nil08}. However, both the aggregate and compound MAC schemes incur significant latency in the verification of the messages because the receiver needs to receive and process all associated packets before being able to verify the validity of the MAC.

\para{Challenges.} In the above discussion, we identify three critical challenges in employing MACs for IoT networks: (1)~incurring minimal communication overhead so that the MAC can fit in a packet, (2)~ensuring that the cryptographic strength meets the security need of the application, and (3)~incurring minimal latency so that the MAC generation and verification processes do not cause unacceptable delays in the packet processing.

\para{Proposed Solution.} In this paper, we addresses the aforementioned challenges through a novel approach for message authentication that we refer to as \textit{Cumulative Message Authentication Code} (CuMAC). In CuMAC, a sender utilizes a procedure called \textit{aggregation} through which the sender first divides the full-sized MAC of each message into multiple short MAC segments, and then ``aggregates'' the MAC segments of multiple messages using a systematic encoding procedure to form a short authentication tag. This procedure resolves the first challenge of ensuring low communication overhead. 

Further, the receiver utilizes a procedure called \textit{accumulation} through which it  first verifies the MAC segments aggregated into the authentication tag of each received packet, and then ``accumulates'' the cryptographic strength by collecting the verified MAC segments associated with the target message. In this procedure, the receiver may incur delay that is proportional to the accumulated cryptographic strength since it needs to wait for the relevant tags to be received and processed. Hence, while the accumulation procedure caters to the second and the third challenge, it brings up a novel and flexible trade-off between the cryptographic strength and latency. CuMAC enables the receiver to authenticate the message in real-time with the cryptographic strength which is commensurate with the size of each tag. Meanwhile, CuMAC also enables the authentication with the highest level of cryptographic strength after accumulating all segments of the MAC that covers the message in the associated packets.

Moreover, in latency-sensitive IoT applications, the receiver may be required to immediately authenticate a message with high cryptographic strength as it arrives. In such cases, the trade-off made by CuMAC may not be sufficient. To address this need, we propose a variant of CuMAC called \textit{CuMAC with Speculation} (\emph{CuMAC/S}) that enables a receiver to accumulate the MAC's cryptographic strength while incurring minimal delay. The core concept of CuMAC/S is motivated by the technique of \emph{speculative execution}\footnote{Speculative execution is an optimization technique in which a computer system performs speculative execution where some outcome is predicted, and execution proceeds along a predicted path. Work is done before it is known whether it is actually needed, so as to prevent a delay that would have to be incurred by doing the work after it is known that it is needed.} which is widely employed in modern computer systems \cite{Cha99,Nig05}. CuMAC/S can be utilized in IoT applications where future messages can be predicted correctly with high reliability with an appropriate speculation model using the current and past messages.

In CuMAC/S, a sender speculates future messages, computes the corresponding MACs, and aggregates the MAC segments of the speculated messages into the authentication tag of the current packet. If the speculated value of a received message is equal to the actual value, then all its segments can be verified in current and previous tags, and hence the receiver can \emph{accumulate} cryptographic strength without having to wait for tags included in forthcoming packets; this significantly cuts down on the MAC verification delay.

The paper's main contributions are summarized as follows.

\begin{itemize}

\item We propose a novel message authentication scheme called \emph{CuMAC}, which meets the security need of resource-constrained IoT applications. CuMAC is an embodiment of two novel concepts that we refer to as \emph{aggregation} (which reduces the communication overhead) and \emph{accumulation} (which increases the cryptographic strength).

\item We propose a variant of CuMAC called CuMAC/S that meets the security need of delay-sensitive, resource-constrained IoT applications. CuMAC/S enables accumulation of cryptographic strength while incurring minimal delay by employing the novel idea of speculation.  

\item We have thoroughly evaluated the effectiveness of CuMAC and CuMAC/S through a simulated in-vehicle controller area network and a prototype implementation on a real car. Our results illustrate that while incurring the same communication overhead as the truncated MAC scheme, CuMAC achieves the cryptographic strength equivalent to the conventional MAC scheme at the cost of increased latency. Further, for the messages which can be accurately speculated, CuMAC/S achieves the cryptographic strength equivalent to the conventional MAC scheme without any additional latency.


\end{itemize}


\section{Motivation for Short MACs}
\label{sec:related}

\begin{figure}[t]
  \centering
  \includegraphics[width=\linewidth]{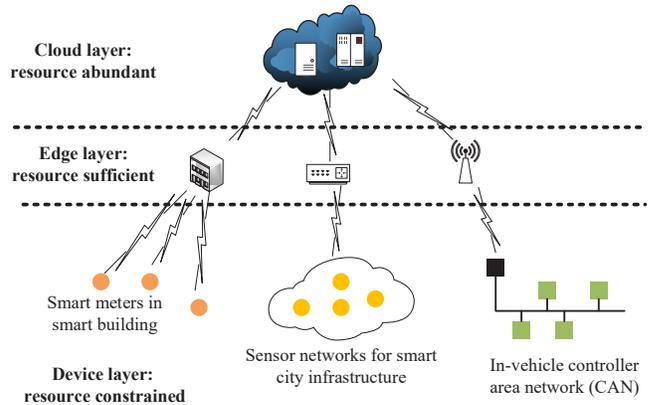}
  \caption{Architecture of typical IoT networks.}
  \savesp
  \label{fig:smart_city}
\end{figure}

IoT networks consist of resource-constrained devices at the lowest layer as shown in Figure \ref{fig:smart_city}. To enable message authentication in such networks, it is imperative to use short MACs as demonstrated by the following discussion of two specific application scenarios -- one with the energy-constrained devices and another with the bandwidth-constrained devices. 

\subsection{Low-Power Wide-Area Network {(LPWAN)}}
\label{sec:applications:LPWAN}

Many IoT applications (e.g., smart metering and smart city infrastructure) require a densely deployed network of low-cost energy-constrained battery-operated wireless devices. The paradigm of LPWAN is aimed at fulfilling these requirements of IoT networks \cite{Raz17,Wan17}.   
Sigfox \cite{Sig17} is one example of a widely-known LPWAN technology. In Sigfox, each uplink packet contains a counter, a message (with length between 0 and 96 bits), and an authentication tag (with length between 16 and 40 bits). To enable robust communication over the unreliable wireless channel, the sender in Sigfox transmits multiple copies of the same packet sequentially. After transmitting the fixed number of copies of the packet, Sigfox waits for an acknowledgement from the receiver. In the absence of the acknowledgement, the packet is considered lost. We note that Sigfox does not support retransmission of lost packets. 

\begin{figure}[t]
    \centering
    \includegraphics[width=\linewidth]{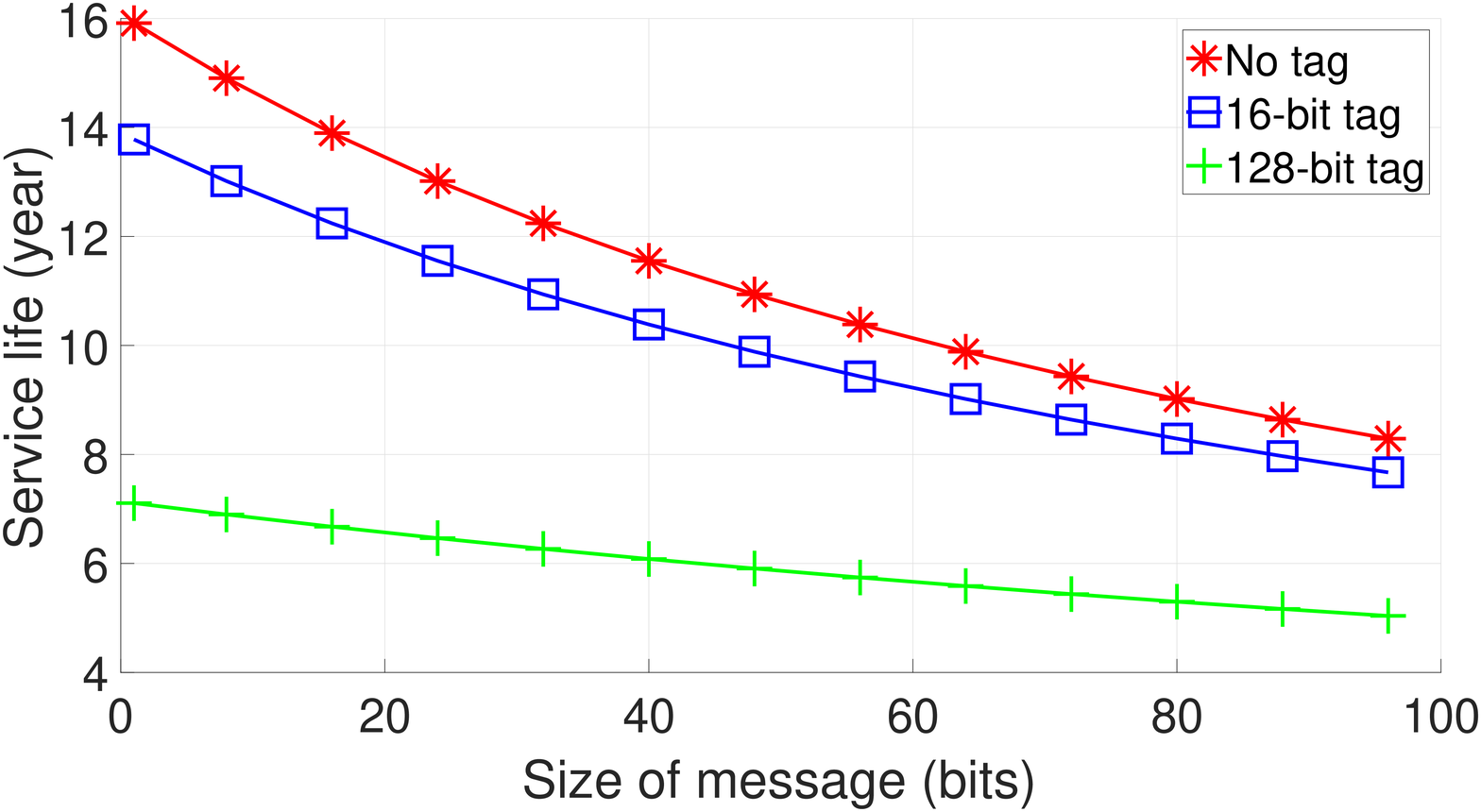}
    \caption{Effect of the size of message and authentication tag on the service life of a sensor node in Sigfox (the results are obtained using the battery consumption data from a Sigfox compliant transceiver produced by ON Semiconductor \cite{On18}.)}
    \savesp
    \label{fig:Life_vs_MacLen}
\end{figure}

The battery-powered Sigfox devices are expected to have a service/battery life of several years. As the energy consumption of a Sigfox device is directly proportional to the size of communicated packets, it is imperative to communicate using short packets to ensure a long service life. Figure~\ref{fig:Life_vs_MacLen} illustrates that in comparison to the standard benchmark of 48-bit messages without any tags, while utilizing a short MAC with 16-bit tags achieves modest (around 10\%) reduction in the service life, utilizing the conventional MAC with 128-bit tags results in a significant loss of around 45\% of the service life. As such, although the message integrity and authentication are of prime importance in applications supported by Sigfox \cite{rom13}, the energy overhead of communicating the full-sized MAC output in the Sigfox packet is undesirably high.

\subsection{In-Vehicle Controller Area Network {(CAN)}}
\label{sec:applications:CAN}

\begin{figure}[t]
  \centering
  \includegraphics[width=0.9\linewidth]{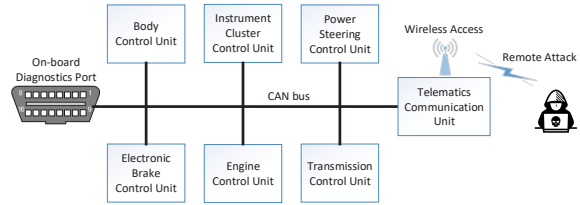}
  \caption{Architecture of an in-vehicle controller area network.}
  \label{fig:CAN_network}
  \savesp
\end{figure}

Today's high-end cars use a hundred or more electronic control units (ECUs) to enable advanced functionalities, such as adaptive cruise control. As shown in Figure \ref{fig:CAN_network}, these ECUs communicate with each other over a bandwidth-constrained wired broadcast channel called the CAN bus \cite{Bos91,ISO15}.
Because the messages communicated among ECUs directly affect vital functions of a vehicle, some of which are safety related (e.g., dynamics control system \cite{Joh05}), the security and reliability of the CAN bus and the integrity of the messages on it are critical~\cite{Soj14}. We note that while the state-of-the-art CAN bus supports robust mechanisms for message acknowledgement and retransmission of corrupted/lost packets, it does not support any security mechanism~\cite{Zag18}. Several studies have shown that a car's in-vehicle network can be compromised through either direct physical access (e.g., using the on-board diagnostics port) or a remote connection (e.g., using Bluetooth) to the CAN bus \cite{Che11,Kos10}. Due to one such vulnerability, Jeep had to recall 1.4 million vehicles in 2015 \cite{Mil15}. To counter such attacks and protect the messages on the CAN bus, the US National Highway Traffic Safety Administration (NHTSA) recommends the inclusion of MACs \cite{Nat16}.

A CAN packet consists of an 11-bit or a 29-bit identifier field and a message field with length between 0 and 64 bits. Except the identifier and message fields, we cannot arbitrarily change the length or the content of other fields in the CAN packet as that would make the modified packet incompatible with the existing CAN protocol. Hence, in the prior art \cite{Szi08,Ued15}, to realize MAC-based authentication in each packet, the identifier field is used to accommodate an 18-bit counter, and the message field is used to accommodate the message payload as well as the authentication tag. Although such a design of the modified packet ensures that it is backward-compatible, inserting a full-sized MAC in the modified packet is not possible because the maximum allowed length of the message field in a CAN packet is only 64 bits.

\para{Proposed Design.} In both the above application scenarios (LPWAN and CAN), the constraints of the IoT network -- either in terms of the MAC size or energy/bandwidth consumption of the networked devices -- prohibit the use of the conventional MAC scheme. To address this challenge, we propose CuMAC and CuMAC/S which can be readily employed in these scenarios to achieve the desired level of security provided by short MACs. In this paper, we utilize CAN as a concrete application scenario to highlight the advantages of CuMAC and CuMAC/S. However, these two schemes can be applied to other resource-constrained network applications, including IoT applications (e.g. LPWAN, Bluetooth Low Energy (BLE)~\cite{gomez2012overview}, Constrained Access Protocol (CoAP)~\cite{bormann2012coap} or Message Queue Telemetry Transport (MQTT)~\cite{singh2015secure}).

\section{Model and Security Objectives}
\label{sec:model}


Here we discuss the network model and define the security objectives of the proposed MAC schemes.

\subsection{Model and Assumptions}

\para{System Model.}
We consider an energy-constrained and/or bandwidth-constrained IoT network where a sender needs to transmit security-critical messages to a receiver using small packets. Hence, the sender and the receiver (after sharing a secret key) employ a MAC scheme for message authentication. We let the sender employ a packet format which contains at least three fields: a packet counter, a message, and an authentication tag. We note that these three fields are critical for ensuring any secure message authentication scheme including CuMAC and CuMAC/S. If the network protocol (e.g., Sigfox as discussed in Section~\ref{sec:applications:LPWAN}) employs these fields in the conventional packets by design, we can readily utilize them; otherwise, the packet contents can be modified in the target network protocol (e.g., CAN as discussed in Section~\ref{sec:applications:CAN}) to include these fields. We assume that there exists a message acknowledgement mechanism which enables the sender to know if a particular packet was correctly delivered to the receiver~\cite{Wan11}. The acknowledgement mechanism assisted with the packet counter enables the sender and the receiver to maintain the same sequence of packets. Note that we do not make any assumption about the message retransmission mechanism, i.e., the network may or may not support retransmission.

\para{Threat Model.}
We consider an adversary which aims to forge valid authentication tags for its malicious messages so that it can deceive the authentication scheme at the receiver. While the adversary can eavesdrop the communication channel to obtain packets transmitted by the sender, it does not know the secret key (used for generating and verifying authentication tags) shared between the sender and the receiver. 

\para{Cryptographic Strength.}
We convey the cryptographic strength in bits, where a cryptographic strength of $\lambda$ bits for a scheme means that for any adversary making at most $2^{\lambda}$ queries or taking at most $2^{\lambda}$ time, the probability of successfully launching an attack against the scheme is negligibly small \cite{Ber13}. The cryptographic strength of a conventional MAC depends on three security criteria: (1) the cryptographic strength of the underlying cryptographic primitive, (2) the size and quality of the secret key, and (3) the size of the MAC output. In this paper, we assume that the first and second criteria have been satisfied, and focus only on the third criterion. As such, to achieve a cryptographic strength of $\lambda$ bits, the minimum size of the MAC output (denoted by $L$) should be $\lambda$ bits.

\subsection{Proposed Approach and Security Objectives}
\label{sec:CuMAC:authLevel}

\para{MAC Design.}
In this paper, we present a novel approach to design a MAC scheme, which consists of two distinctive procedures: aggregation and accumulation. In aggregation, the sender generates short authentication tags from segments of multiple MACs. In accumulation, the receiver accumulates the cryptographic strength of the underlying MAC by collecting and verifying the authentication tags. For the scenarios where future messages can be speculated and their corresponding MACs can be pre-computed, the aggregation and accumulation may happen even before the messages are transmitted.

\begin{figure}[t]
    \centering
    \includegraphics[width = \linewidth]{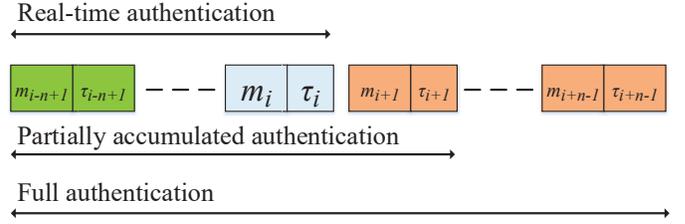}
    \caption{Illustrative distribution of the segments of the MAC of message $m_i$, and definition of the authentication levels.}
    \savesp
    \label{fig:types_of_security}
\end{figure}

We discuss the above approach through the illustration shown in Figure~\ref{fig:types_of_security}. We consider that an $L$-bit MAC of a message $m_i$ is divided into $n$ segments each of length $l$, and distributed in tags $\tau_{i}, \cdots, \tau_{i+n-1}$. Also, if after generating the message $m_{i-n+1}$, the message $m_i$ can be speculated as $\widehat{m}_i$, the corresponding MAC is computed as $\widehat{\sigma}_i$. The MAC $\widehat{\sigma}_i$ is divided into $n$ segments, and the last $n-1$ segments are distributed in tags $\tau_{i-n+1}, \cdots, \tau_{i-1}$, which are transmitted before $\tau_i$.

\para{Authentication Levels.} To compare the proposed approach with prior art, we define three levels/features of authentication: (1)~real-time authentication, (2)~full authentication, and (3)~partially accumulated authentication. Figure~\ref{fig:types_of_security} illustrates these different levels of authentication, when applied to message $m_i$. Here the receiver can perform real-time authentication immediately after receiving message $m_i$ by processing the current tag $\tau_i$ and the previous tags $\tau_{i-n+1}, \cdots, \tau_{i-1}$. With real-time authentication, the receiver performs authentication without any delay, but it achieves the lowest cryptographic strength since there is no security accumulation using the subsequent tags. On the other hand, the receiver can perform full authentication after receiving all of the segments of the MAC associated with message $m_i$ in tags $\tau_{i-n+1}, \cdots, \tau_{i+n-1}$. With full authentication, the receiver achieves the highest cryptographic strength, but needs to incur a latency of $n-1$ packets. The receiver can perform partially accumulated authentication by accumulating and processing tags $\tau_{i-n+1}, \cdots, \tau_{i+r-1}$, where $1 < r < n$. Partially accumulated authentication enables the receiver to make a trade-off between cryptographic strength and message verification latency to meet the security and performance needs of the application.

\para{Security Objectives.}
The security objective of the proposed MAC scheme is to ensure that the probability with which an adversary succeeds in breaking each of the three authentication features is negligible (i.e, as difficult as random guessing). 
Specifically, to break the real-time authentication feature, the adversary needs to forge a message and a valid tag. The forgery need to be \textit{fresh} which means that the sender has not generated the MAC of the same counter and message pair using the same shared key. As such, the cryptographic strength of real-time authentication is limited by the size of the MAC segment $l$.
To break the partially accumulated authentication feature with $r$ accumulated segments, the adversary need to forge a sequence of $r$ messages with valid tags. In this sequence, the forgery for only the first message needs to be fresh. Hence, the cryptographic strength of partially accumulated authentication depends on the size of the MAC segment $l$ and the number of accumulated segments $r$. Similarly, to break the full authentication feature, the adversary needs to forge a sequence of $n$ messages with valid tags, where forgery for at least the first message is fresh. Hence, the cryptographic strength of full authentication is limited by the size of MAC $L$. In the case of the speculation of future messages, the cryptographic strengths of the real-time and the partially accumulated authentication also depend on the message speculation accuracy. 

A formal discussion of the security properties and associated proofs corresponding to CuMAC and CuMAC/S are provided appendix \ref{sec:security_definition} and appendix \ref{sec:security_proof}. 

\section{Technical Details of CuMAC}
\label{sec:CuMAC}

\begin{figure}[t]
    \centering
    \includegraphics[width = \linewidth]{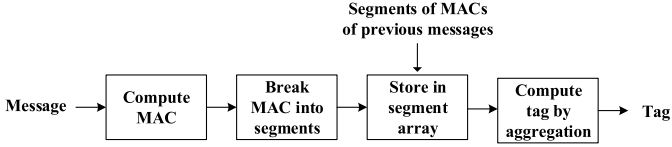}
    \caption{Schematic of the procedures at the sender in CuMAC.}
    \savesp
    \label{fig:CuMAC_block}
\end{figure}

CuMAC comprises of two major algorithms: tag generation and tag verification. In the tag generation algorithm, the sender computes the authentication tag through two major steps (Figure~\ref{fig:CuMAC_block}). In the first step, the sender generates the MAC of the message, breaks the MAC into short segments, and stores them into a segment array. In the second step, the sender retrieves one MAC segment of the current message, and several segments of the MACs of the previously transmitted messages from the segment array, and aggregates the segments to generate a tag. Having received each packet, the receiver runs the tag verification algorithm which includes two major steps. In the first step, the receiver generates an authentication tag of the received message using the same procedure employed in the tag generation algorithm. In the second step, the receiver compares the generated authentication tag with the received authentication tag. If the authentication tags match, the receiver accumulates the MAC segment (aggregated in the authentication tag) with the previously received MAC segments of the corresponding message. 

Below we present the technical details of the algorithms in CuMAC, and an instantiation that illustrates the generation and verification of the tags in CuMAC.

\subsection{Algorithms}

CuMAC is composed of the following algorithms.

\smallskip

\noindent
$\mathtt{k} \gets \bm{\mathsf{KeyGen}}(1^\lambda)$

This probabilistic key generation algorithm is utilized by the sender and receiver to obtain the secret key. The input to this algorithm is the security parameter $\lambda \in \NN$, and the output is the secret key denoted by $\mathtt{k}$. In a resource-constrained network, this algorithm can be efficiently realized by leveraging a trusted third party \cite{sch11}. In the absence of such a party, it can also be realized using an efficient key distribution scheme~\cite{du05}.

\smallskip

\noindent
$\sigma_i \gets \bm{\mathsf{MacGen}}(\mathtt{k},i,m_i)$

This deterministic MAC generation algorithm is utilized by the sender and the receiver (as a sub-algorithm of tag generation and verification algorithms) to compute the MAC of a message using the secret key. The inputs to this algorithm are the secret key $\mathtt{k}$, a counter $i$ and a message $m_i$. This algorithm outputs the $L$ bits long MAC represented by $\sigma_i$. This algorithm can be realized using a cipher-based (e.g., AES-CMAC~\cite{Bel00}) or a hash-based (e.g., SHA-3) MAC scheme. In this paper, we utilize the widely used AES-CMAC. 

\smallskip

\noindent 
$\tau_i \gets \bm{\mathsf{SegAgg}}(\mathtt{segArray})$

This segment aggregation algorithm is utilized by the sender and the receiver as a sub-algorithm of tag generation and tag verification algorithms, respectively. It takes as input a two-dimensional array of MAC segments $\mathtt{segArray}$. This algorithm proceeds as follows. The $i^{\mathrm{th}}$ row of segments in $\mathtt{segArray}$ is generated as follows. The $L$-bit MAC $\sigma_i$ is divided into $n$ segments, such that the size of each segment is $l$ bits, i.e., $L=n \cdot l$. The $j^{\mathrm{th}}$ segment of $\sigma_i$ is represented by $s_i^j$, and is extracted from $\sigma_i$ as 
\begin{align}
\label{eq:segment}
s_i^j \gets (\sigma_i)_{\downarrow [(j-1) \cdot l+1,j \cdot l]}.
\end{align}
It means that the bits in $s_i^j$ correspond to the bits from $\left((j-1) \cdot l+1\right)^{\mathrm{th}}$ bit to $\left(j \cdot l\right)^{\mathrm{th}}$ bit in $\sigma_i$. Further, this algorithm extracts $n$ elements from $\mathtt{segArray}$ ($n-1$ previous MAC segments and one current MAC segment), and computes the authentication tag $\tau_i$ as follows.
\begin{align}
\label{eq:tag}
\tau_i \gets \bigoplus_{j=1,i-j+1>0}^{n} s_{i-j+1}^{j}.
\end{align}
This algorithm outputs the authentication tag $\tau_i$.

\smallskip

\noindent
$\tau_i \gets \bm{\mathsf{TagGen}}(\mathtt{k},i,m_i)$ 

This tag generation algorithm is run by the sender to generate an authentication tag. It takes as inputs the secret key $\mathtt{k}$, a counter $i$ and a message $m_i$. It utilizes an array of MAC segments $\mathtt{segTx}$ which is stored and maintained by the sender. This algorithm proceeds as follows to output the authentication tag $\tau_i$.

\begin{enumerate}

\item Compute the MAC of the message $m_i$ and set it as $\sigma_i$, i.e., $\sigma_i \gets \mathsf{MacGen}(\mathtt{k},i,m_i)$. 

\item Divide the MAC $\sigma_i$ into $n$ segments as shown in equation~\eqref{eq:segment} and append the segments to the array $\mathtt{segTx}$.

\item Compute and output the tag $\tau_i$ by aggregating the segments of MACs in $\mathtt{segTx}$ as shown in equation~\eqref{eq:tag}, i.e., $\tau_i \gets \mathsf{SegAgg}(\mathtt{segTx})$. 
\end{enumerate}  

After receiving a positive acknowledgment of the delivery of the packet at the receiver, the sender increments the counter~$i$ by one for the next packet. We note that the counter~$i$ can be readily employed to handle the case of a lost packet. The sender gets to know that the $i^{\mathrm{th}}$ packet is lost when it does not receive the acknowledgement from the receiver or it receives a negative acknowledgement. In this case, if the sender does not support any retransmission mechanism, the sender does not increment the packet counter, removes the $i^{\mathrm{th}}$ row (i.e., the most recently appended row) of segments in $\mathtt{segTx}$, and then proceeds with the tag generation of the next message.

\smallskip

\noindent
$\mathtt{valid}/\mathtt{invalid} \gets \bm{\mathsf{TagVerify}}(\mathtt{k},i,m_i,\tau_i)$ 
 
This verification algorithm is run by the receiver for verifying the authenticity of the received message and tag. It takes as inputs the secret key $\mathtt{k}$, the received counter $i$, the received message $m_i$, and the received tag $\tau_i$. It also utilizes an array of MAC segments $\mathtt{segRx}$ and an array of verified MAC segments $\mathtt{accRx}$. These arrays are stored and maintained by the receiver. This algorithm first generates the tag for the received message using the $\mathsf{TagGen}$ algorithm while updating the MAC segments in $\mathtt{segRx}$, i.e., $\widetilde{\tau}_i \gets \mathsf{TagGen}(\mathtt{k},i,m_i)$. It then verifies whether the generated tag $\widetilde{\tau}_i$ is equal to the received tag $\tau_i$. If the verification succeeds, it updates the array of accumulated MAC segments $\mathtt{accRx}$ and outputs the value $\mathtt{valid}$; otherwise, it outputs the value $\mathtt{invalid}$.

\begin{table}[t]
\centering
\caption{Example illustrating CuMAC with $L=128$, $n = 4$, and $l = 32$.}
\label{tab:example_CuMAC}
\scalebox{0.85}{
\renewcommand{\arraystretch}{1.2}
\begin{tabular}{|c|c|c|l|c|}
\hline
\textbf{Packet} & \textbf{Previous} & \textbf{Current} & \multirow{2}{*}{\textbf{Aggregation of MAC segments}}  & \multirow{2}{*}{\textbf{Tag}} \\

\textbf{Counter} & \textbf{MACs} & \textbf{MAC}  &  &  \\
\hline

5 & $\sigma_2,\sigma_3,\sigma_4$ & $\sigma_5$  &  $s_2^{4} \oplus s_3^{3} \oplus s_4^{2} \oplus s_5^{1}  $ & $\tau_5$\\ 
\hline

6 & $\sigma_3,\sigma_4,\sigma_5$ &$\sigma_6$ & \hspace{15pt} $s_3^{4} \oplus s_4^{3} \oplus s_5^{2} \oplus s_6^{1}  $  & $\tau_6$\\ 
\hline

7 & $\sigma_4,\sigma_5,\sigma_6$ &$\sigma_7$ & \hspace{33pt} $s_4^{4} \oplus s_5^{3} \oplus s_6^{2} \oplus s_7^{1}$  & $\tau_7$\\ 
\hline

8 & $\sigma_5,\sigma_6,\sigma_7$ &$\sigma_8$ & \hspace{51pt} $s_5^{4} \oplus s_6^{3} \oplus s_7^{2} \oplus s_8^{1} $ & $\tau_8$\\ 
\hline

\end{tabular}}
\end{table}

\subsection{Illustration}
\label{sec:cumac:example}

Table~\ref{tab:example_CuMAC} presents an example of CuMAC. The size of the tag in each packet is $32$ bits (i.e., $l = 32$). The MAC is generated using the AES-CMAC algorithm. Hence, the size of the MAC output is $128$~bits (i.e., $L=128$), which provides cryptographic strength of $128$ bits. Each MAC is divided into four segments (i.e., $n = 4$). In the fifth packet, the MAC $\sigma_5$ of the message $m_5$ is computed. To compute the corresponding tag $\tau_5$, the sender aggregates the segment $s_5^{1}$ of the MAC $\sigma_5$ and the segments of the MACs of the previously generated messages, $\sigma_2$, $\sigma_3$ and $\sigma_3$. Further, the tags $\tau_6$, $\tau_7$ and $\tau_8$ are computed using the segments $s_5^{2}$, $s_5^{3}$ and $s_5^{4}$ of $\sigma_5$, respectively.

When the receiver receives the fifth packet with the message $m_5$, the successful verification of the tag $\tau_5$ enables the real-time authentication of message $m_5$ with the cryptographic strength of $32$~bits. Next, the receiver receives and verifies the validity of tags $\tau_6$, $\tau_7$, and $\tau_8$. If all four tags are verified as $\mathtt{valid}$, the receiver combines the segments $s_5^{1}$, $s_5^{2}$, $s_5^{3}$ and $s_5^{4}$---which are contained in tags $\tau_5$, $\tau_6$, $\tau_7$ and $\tau_8$, respectively---to accumulate the cryptographic strength. This enables the receiver to perform full authentication of message $m_5$ with the cryptographic strength of $128$ $(=4 \times 32)$ bits. However, if the receiver is restricted to process the fifth packet only after receiving the seventh packet due to latency requirements, it may also perform partially accumulated authentication of message $m_5$ with a cryptographic strength of $96$~bits after verifying tags $\tau_5$, $\tau_6$ and $\tau_7$. We highlight that this ability to perform the partially accumulated authentication is the most unique feature of CuMAC when compared to prior art.
\section{Technical Details of CuMAC/S}
\label{sec:CuMAC/S}

In latency-sensitive applications, the receiver must authenticate a message as it arrives. For such applications, the trade-off between the cryptographic strength and latency made by CuMAC may not be sufficient. To address this challenge, we present CuMAC/S, which employs a novel concept of message speculation for MAC generation. Equipped with an accurate message speculation algorithm, CuMAC/S achieves both high cryptographic strength and low verification latency.

\subsection{Feasibility of Speculation}
\label{subsec:accuracy}

\begin{figure}
    \begin{subfigure}[b]{0.49\linewidth}
    \includegraphics[width=\textwidth]{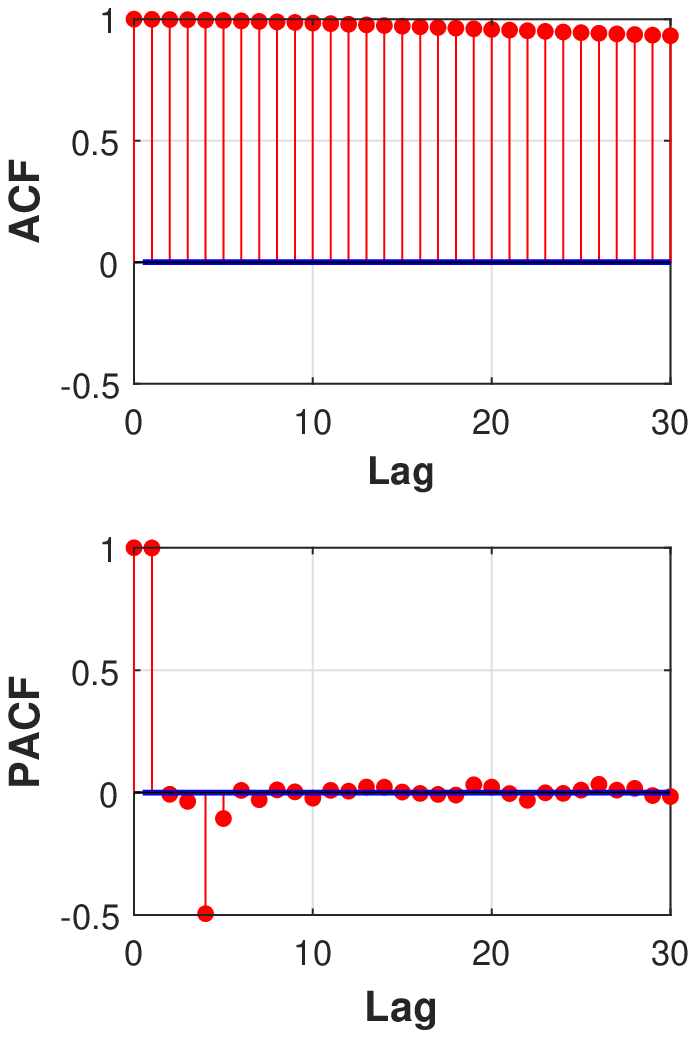}
    \caption{Original data.}
    \label{fig:arima_d0}
    \end{subfigure}
    \begin{subfigure}[b]{0.49\linewidth}
    \includegraphics[width=\textwidth]{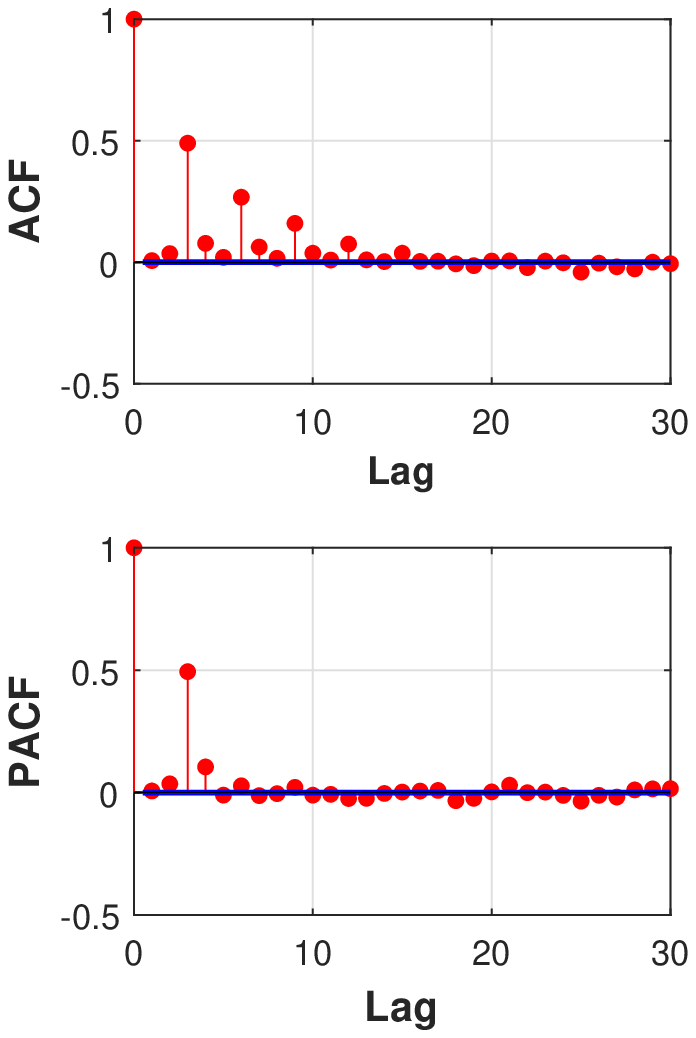}
    \caption{First-order differenced data.}
    \label{fig:arima_d1}
    \end{subfigure}
    \caption{Example illustrating the feasibility of speculation of a vehicle's transmission torque values through an ARIMA model whose parameters are determined using the autocorrelation function (ACF) and partial autocorrelation function (PACF).}
    \savesp
    \label{fig:arima}
\end{figure}

We discuss the feasibility of speculation of future messages by analyzing the messages communicated on a CAN bus in a typical vehicle. To evaluate the speculation accuracy for different CAN messages, we utilize trace files of a real vehicle, which have been recorded using the OpenXC platform \cite{FordOpenXC}. These files present different types of CAN messages which can be identified and interpreted by OpenXC libraries. To speculate future message values, we utilize autoregressive integrated moving average (ARIMA) model, which is a widely used model for time series analysis. 

The ARIMA model with hyper parameters $(p,d,q)$ implies that for the $d^{\mathrm{th}}$-order difference of the time series values, a speculated future message value is the linear combination of $p$ previous values, and $q$ previous error values in the speculation. We utilize the Box-Jenkins~\cite{mak97} method to compute the hyper parameters of the ARIMA model for each type of CAN messages. In this method, we determine the values for $p$, $d$ and $q$ by observing the autocorrelation and partial autocorrelation of the message values. We illustrate this procedure in Figure~\ref{fig:arima} which presents the autocorrelation and partial autocorrelation of the values of the message corresponding to the torque at transmission in a vehicle. From the results shown in Figure~\ref{fig:arima_d0}, we observe that there is high autocorrelation between message values. Further, from the results shown in Figure~\ref{fig:arima_d1}, we observe that the autocorrelation decays gradually, and the partial autocorrelation is close to zero after a lag of $3$ message values. Hence, according to the rules of Box-Jenkins approach, we set ARIMA(3,1,0) model to speculate the message values corresponding to the torque at transmission.

In our analysis, we train the ARIMA model using the first 90$\%$ of the message values, and then we employ the model on the last 10$\%$ of message values for the test. Here, the accuracy of correct speculation/prediction of future message values is measured using a metric called speculation error rate (SER), which is defined as the ratio between the number of incorrect speculations and the total number of speculations. Note that the speculation is correct only if the message is correctly predicted up to the least significant bit (LSB). Table~\ref{tab:arima} shows the speculation error rate of ten message-types. We observe that certain types of CAN messages (e.g. the first five message types listed in Table~\ref{tab:arima}) can be predicted with high reliability using the ARIMA model.

We can improve the speculation accuracy by using more sophisticated and further tuned models. Moreover, we can mitigate the impact of speculation errors by increasing robustness of the MAC scheme against such errors. For example, if the message contains some values for which some of the least significant bits can be safely ignored (without impacting performance or security), then these bits do not need to be protected by a MAC, and hence the MAC calculation can be limited  to only the part of a message that can be predicted with high reliability. In the rightmost column of Table~\ref{tab:arima}, we show that the SER can be significantly improved for some types of messages by ignoring the last three least LSBs.

Now we present the technical details of the algorithms in CuMAC/S, and an instantiation that illustrates the generation and verification of the tags in CuMAC/S.  

\begin{table}[t]
\centering
\caption{Speculation accuracy for typical CAN messages.}
\label{tab:arima}
\renewcommand{\arraystretch}{1.3}
\scalebox{0.95}{
\begin{tabular}{|l|c|c|}
\hline
\textbf{Signal}                    & \textbf{SER}       & \textbf{SER after ignoring 3 LSBs} \\ \hline
Longitude                          & $<$0.0001                  & $<$0.0001                  \\ \hline
Latitude                           & $<$0.0001                  & $<$0.0001                  \\ \hline
Odometer                           & $<$0.0001                 & $<$0.0001                  \\ \hline
Fuel level                         & $<$0.0001                  & $<$0.0001                  \\ \hline
Fuel consumed since restart        & $<$0.0001                  & $<$0.0001                  \\ \hline
Accelerator pedal position         & 0.0030             & 0.0002             \\ \hline
Torque at transmission             & 0.0100             & 0.0020             \\ \hline
Engine speed                       & 0.2329             & 0.0880             \\ \hline
Vehicle speed                      & 0.2478             & 0.0975             \\ \hline
Steering wheel angle               & 0.4763             & 0.3595             \\ \hline
\end{tabular}}
\end{table}

\begin{table*}[t]
\centering
\caption{Example illustrating CuMAC/S with $L=128$, $n = 4$, and $l = 32$.}
\label{tab:example_CuMAC/S}
\scalebox{1}{
\renewcommand{\arraystretch}{1.2}
\begin{tabular}{|c|c|c|c|c|l|c|}
\hline
\textbf{Packet} & \textbf{Previous} & \textbf{Current} & \textbf{Previous} & \textbf{Current} & \hspace{40pt} \multirow{2}{*}{\textbf{Aggregation of MAC segments}}  & \multirow{2}{*}{\textbf{Tag}} \\

\textbf{Counter} & \textbf{MACs} & \textbf{MAC} & \textbf{speculated MACs} & \textbf{speculated MAC} &  &  \\
\hline 
2 & $ \sigma_1$ & $\sigma_2$ & $\widehat{\sigma}_3, \widehat{\sigma}_4$ & $\widehat{\sigma}_5$ & $  s_1^2 \oplus s_2^1 \oplus \widehat{s}_3^2 \oplus \widehat{s}_4^3 \oplus \widehat{s}_5^4$ & $\tau_2$\\ 
\hline

3 & $\sigma_1,\sigma_2$ & $\sigma_3$ & $\widehat{\sigma}_4, \widehat{\sigma}_5$ & $\widehat{\sigma}_6$ &  $ s_1^3 \oplus s_2^2 \oplus s_3^1 \oplus \widehat{s}_4^2 \oplus \widehat{s}_5^3 \oplus \widehat{s}_6^4$ & $\tau_3$\\ 
\hline

4 & $\sigma_1,\sigma_2,\sigma_3$ & $\sigma_4$ & $\widehat{\sigma}_5,\widehat{\sigma}_6$ & $\widehat{\sigma}_7$ &  $s_1^4 \oplus s_2^3 \oplus s_3^2 \oplus s_4^1 \oplus \widehat{s}_5^2 \oplus \widehat{s}_6^3 \oplus \widehat{s}_7^4 $ & $\tau_4$\\ 
\hline

5 & $\sigma_2,\sigma_3,\sigma_4$ &$\sigma_5$& $\widehat{\sigma}_6, \widehat{\sigma}_7$ & $\widehat{\sigma}_8$ & \hspace{15pt} $s_2^4 \oplus s_3^3 \oplus s_4^2 \oplus s_5^1 \oplus \widehat{s}_6^2 \oplus \widehat{s}_7^3 \oplus \widehat{s}_8^4 $  & $\tau_5$\\ 
\hline

6 & $\sigma_3,\sigma_4,\sigma_5$ &$\sigma_6$ & $\widehat{\sigma}_7,\widehat{\sigma}_8$ & $\widehat{\sigma}_9$ & \hspace{33pt} $s_3^4 \oplus s_4^3 \oplus s_5^2 \oplus s_6^1 \oplus \widehat{s}_7^2 \oplus \widehat{s}_8^3 \oplus \widehat{s}_9^4$  & $\tau_6$\\ 
\hline

7 & $\sigma_4,\sigma_5,\sigma_6$ &$\sigma_7$ & $\widehat{\sigma}_8,\widehat{\sigma}_9$ & $\widehat{\sigma}_{10}$ & \hspace{51pt} $s_4^4 \oplus s_5^3 \oplus s_6^2 \oplus s_7^1 \oplus \widehat{s}_8^2 \oplus \widehat{s}_9^3 \oplus \widehat{s}_{10}^4$ & $\tau_7$\\ 
\hline

8 & $\sigma_5,\sigma_6,\sigma_7$ &$\sigma_8$ & $\widehat{\sigma}_9,\widehat{\sigma}_{10}$ & $\widehat{\sigma}_{11}$ & \hspace{69pt} $s_5^4 \oplus s_6^3 \oplus s_7^2 \oplus s_8^1 \oplus \widehat{s}_9^2 \oplus \widehat{s}_{10}^3 \oplus \widehat{s}_{11}^4$ & $\tau_8$\\ 
\hline

\end{tabular}}
\end{table*}

\subsection{Algorithms}

The $\mathsf{KeyGen}$ and $\mathsf{MacGen}$ algorithms in CuMAC (discussed in Section~\ref{sec:CuMAC}) and those in CuMAC/S are the same, and hence we do not provide their details in this section. We present the details of other algorithms in CuMAC/S as follows. 

\smallskip

\noindent
$\mathtt{msgArray'} \gets \bm{\mathsf{MsgSpec}}(\mathtt{msgArray})$

This deterministic message speculation algorithm is utilized by the sender and receiver for the speculation of future message values. It takes an array of the transmitted and speculated messages $\mathtt{msgArray}$ as input. At the $i^{\mathrm{th}}$ instance, the array $\mathtt{msgArray}$ can be represented as $\{ m_1, m_2, \cdots , m_{i-1},m_i, \widehat{m}_{i+1}, \widehat{m}_{i+2}, \cdots \widehat{m}_{i+n-2}\}$. This algorithm generates the predicted value of the message $m_{i+n-1}$, which is represented by $\widehat{m}_{i+n-1}$, appends it to the array $\mathtt{msgArray}$, and outputs the updated array $\mathtt{msgArray'}$. Since the speculation model used in this algorithm is deterministic, the sender and the receiver run the same set of steps, and obtain the same speculated messages given the same input messages.

\smallskip

\noindent 
$\tau_i \gets \bm{\mathsf{SegAgg}}(\mathtt{segArray})$

This segment aggregation algorithm is run by the sender and the receiver. It takes as input a two-dimensional array of MAC segments $\mathtt{segArray}$. The $\mathtt{segArray}$ comprises of the segments of the MACs of the transmitted and speculated messages. The $i^{\mathrm{th}}$ entry in $\mathtt{segArray}$ is generated by the segment $s_i^j$ $\forall j \in [1,n]$ using the equation~\eqref{eq:segment}. This algorithm extracts $2n-1$ elements from $\mathtt{segArray}$ ($n-1$ previous MAC segments, current MAC segment, and $n-1$ speculated MAC segments), and computes the authentication tag $\tau_i$ as follows.
\begin{align}
\tau_i \gets \left(\bigoplus_{j=1,i-j+1>0}^{n} s_{i-j+1}^{j}\right) \oplus \left(\bigoplus_{j=2}^{n} \widehat{s}_{i+j-1}^{j}\right).
\end{align}
This algorithm outputs the authentication tag $\tau_i$.

\begin{figure}[t]
    \includegraphics[width = \linewidth]{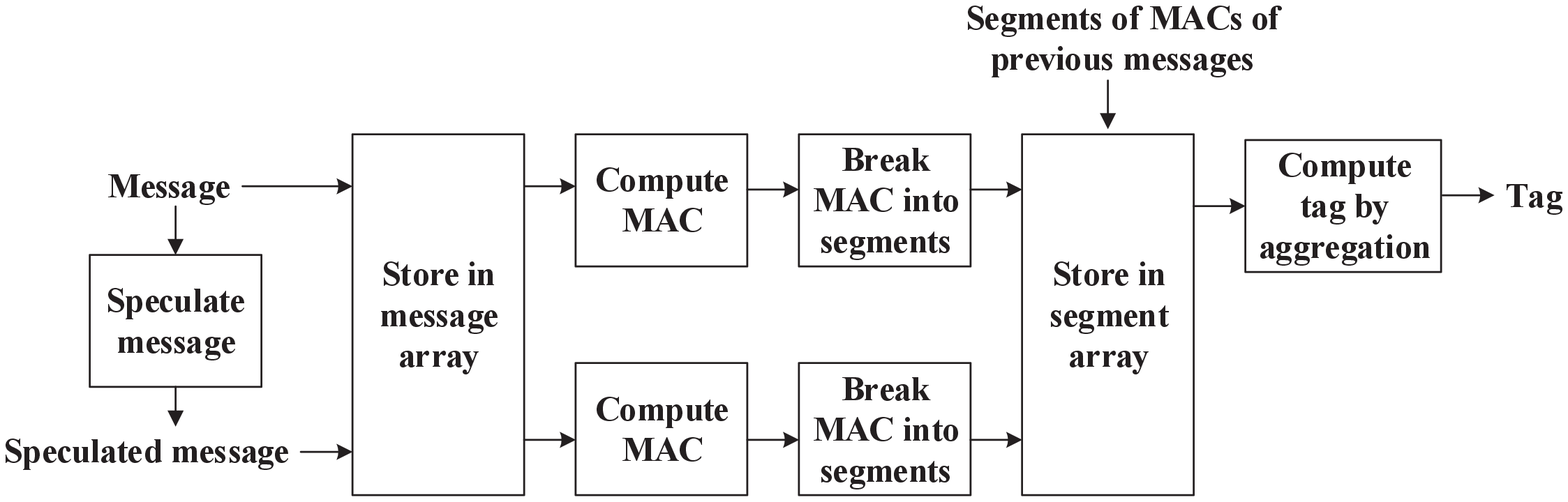}
    \caption{Schematic of the procedures at the sender in CuMAC/S.}
    \savesp
    \label{fig:CuMAC/S_block}
\end{figure} 

\smallskip

\noindent
$\tau_i \gets \bm{\mathsf{TagGen}}(\mathtt{k},i,m_i)$ 

This tag generation algorithm is utilized by the sender to generate an authentication tag. It takes as inputs the secret key $\mathtt{k}$, a counter $i$ and a message $m_i$. It utilizes an array of the transmitted and speculated messages $\mathtt{msgTx}$, and an array of MAC segments of the transmitted and speculated messages $\mathtt{segTx}$. The arrays $\mathtt{msgTx}$ and $\mathtt{segTx}$ are stored and maintained by the sender. Figure~\ref{fig:CuMAC/S_block} presents an overview of the algorithm which proceeds as follows.

\begin{enumerate}
\item Extract $\widehat{m}_i$ from $\mathtt{msgTx}$ and verify whether $m_i = \widehat{m}_i$. 
\begin{enumerate}
\item If $m_i=\widehat{m}_i$, set $\sigma_i=\widehat{\sigma}_i$.
\item Otherwise, if $m_i \neq \widehat{m}_i$, compute the MAC of the message $m_i$ and set it as $\sigma_i$, i.e., $\sigma_i \gets \mathsf{MacGen}(\mathtt{k},i,m_i)$. Divide the MAC $\sigma_i$ into $n$ segments and replace the MAC segments of $\widehat{\sigma}_i$ in the array $\mathtt{segTx}$.
\end{enumerate} 

\item Predict the value of the message $m_{i+n-1}$ and append the speculated message $\widehat{m}_{i+n-1}$ to the array $\mathtt{msgTx}$, i.e.,  $\mathtt{msgTx'} \gets \mathsf{MsgSpec}(\mathtt{msgTx})$.

\item Compute the MAC of the message $\widehat{m}_{i+n-1}$ and set it as $\widehat{\sigma}_{i+n-1}$, i.e., $\widehat{\sigma}_{i+n-1} \gets \mathsf{MacGen}(\mathtt{k},i,\widehat{m}_{i+n-1})$. Divide the MAC $\widehat{\sigma}_{i+n-1}$ into $n$ segments and append to the array $\mathtt{segTx}$.

\item Compute the current tag $\tau_i$ by aggregating the segments of MACs of the previous, current and future messages, i.e., $\tau_i \gets \mathsf{SegAgg}(\mathtt{segTx})$.
\end{enumerate}  

\smallskip

\noindent
$\mathtt{valid}/\mathtt{invalid} \gets \bm{\mathsf{TagVerify}}(\mathtt{k},i,m_i,\tau_i)$ 

This verification algorithm is utilized by the receiver for verifying the authenticity of the received message and tag. It takes as inputs the secret key $\mathtt{k}$, the received counter $i$, the received message $m_i$, and the received tag $\tau_i$. It stores and manages an array of previously received and speculated messages $\mathtt{msgRx}$, an array of MAC segments $\mathtt{segRx}$, and an array of verified segments $\mathtt{accRx}$. This algorithm first generates the tag for the received message using the $\mathsf{TagGen}$ algorithm while updating the speculated message in $\mathtt{msgRx}$ and MAC segments in $\mathtt{segRx}$, i.e., $\widetilde{\tau}_i \gets \mathsf{TagGen}(\mathtt{k},i,m_i)$. It then verifies whether the generated tag $\widetilde{\tau}_i$ is equal to the received tag $\tau_i$. If the verification succeeds, it updates the array of accumulated MAC segments $\mathtt{accRx}$ and outputs the value $\mathtt{valid}$; otherwise, it outputs the value $\mathtt{invalid}$. 

\subsection{Illustration}
\label{sec:cumac-s:example}
Table~\ref{tab:example_CuMAC/S} presents an example of CuMAC/S, which follows the example of CuMAC presented in Section~\ref{sec:cumac:example}. When the receiver receives the fifth packet with the message $m_5$, it verifies whether it matches the speculated message $\widehat{m}_5$. If they do not match, the successful verification of the tag $\tau_5$ enables the real-time authentication of message $m_5$ with a cryptographic strength of $32$~bits, which is the same as in CuMAC. Next, the receiver receives and verifies the validity of tags $\tau_5, \cdots, \tau_8$. If all four tags are verified as $\mathtt{valid}$, the receiver combines the segments $s_5^1$, $s_5^2$, $s_5^3$ and $s_5^4$---which are contained in tags $\tau_5$, $\tau_6$, $\tau_7$ and $\tau_8$, respectively---to accumulate the cryptographic strength. This enables the receiver to perform full authentication of message $m_5$ with a cryptographic strength of $128$ $(=4 \times 32)$ bits.

However, if the message $m_5$ and $\widehat{m}_5$ match, and tags $\tau_2$, $\tau_3$, $\tau_4$ and $\tau_5$ are verified as $\mathtt{valid}$, the receiver combines the segments $s_5^1$, $\widehat{s}_5^2$, $\widehat{s}_5^3$ and $\widehat{s}_5^4$---which are contained in tags $\tau_5$, $\tau_4$, $\tau_3$ and $\tau_2$, respectively. This enables the receiver to perform real-time authentication of message $m_5$ with a cryptographic strength of $128$ bits. We highlight that this unique ability to achieve equal cryptographic strengths for the real-time authentication and full authentication in spite of using short authentication tags distinguishes CuMAC/S from prior art.
\section{Simulation Results}
\label{sec:evaluation}

In this section, we consider a simulated IoT environment, and evaluate the performance of CuMAC and CuMAC/S by comparing them with three other schemes from the prior art: the truncated MAC \cite{Szi08}, the compound MAC \cite{Nil08}, and the aggregate MAC \cite{Kat08}. For all five schemes, AES-CMAC with a MAC output of 128~bits is utilized as the underlying MAC algorithm. We set the size of the tag in all schemes to 16~bits. In the truncated MAC scheme, each MAC is truncated to 16~bits, and transmitted as the tag. In the compound MAC scheme, a compound MAC of 128 bits is computed over eight messages. In the aggregate MAC scheme, an aggregate MAC of 128~bits is computed by aggregating the MACs of eight messages. The compound MAC and the aggregate MAC are divided into eight segments each of size 16~bits, and transmitted in each of the eight packets as the tag. In CuMAC and CuMAC/S, each MAC of 128~bits is divided into eight segments each of size 16~bits. In CuMAC, each tag is generated by aggregating segments of seven previously transmitted messages and the current message. In CuMAC/S, each tag is generated by aggregating segments of seven previously transmitted messages, the current message, and seven speculated messages.

\begin{figure}[t]
\centering
\begin{subfigure}[t]{0.48\linewidth}
    \centering
    \includegraphics[width=\linewidth]{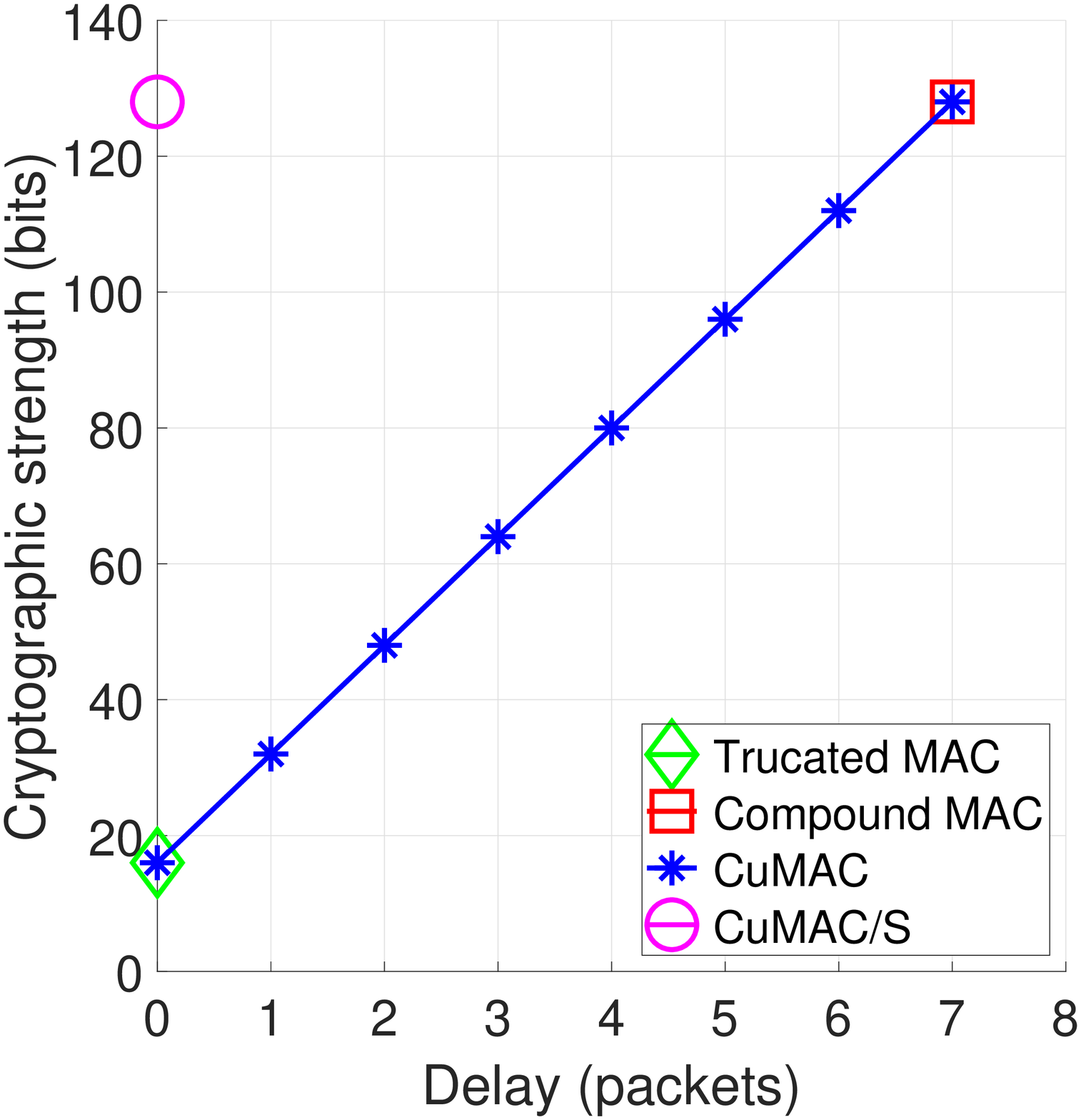}
    \caption{Trade-off between cryptographic strength and delay.}
    \label{fig:accumulating-security_vs_delay}
\end{subfigure}
\hfill
\begin{subfigure}[t]{0.48\linewidth}
    \centering
    \includegraphics[width=\linewidth]{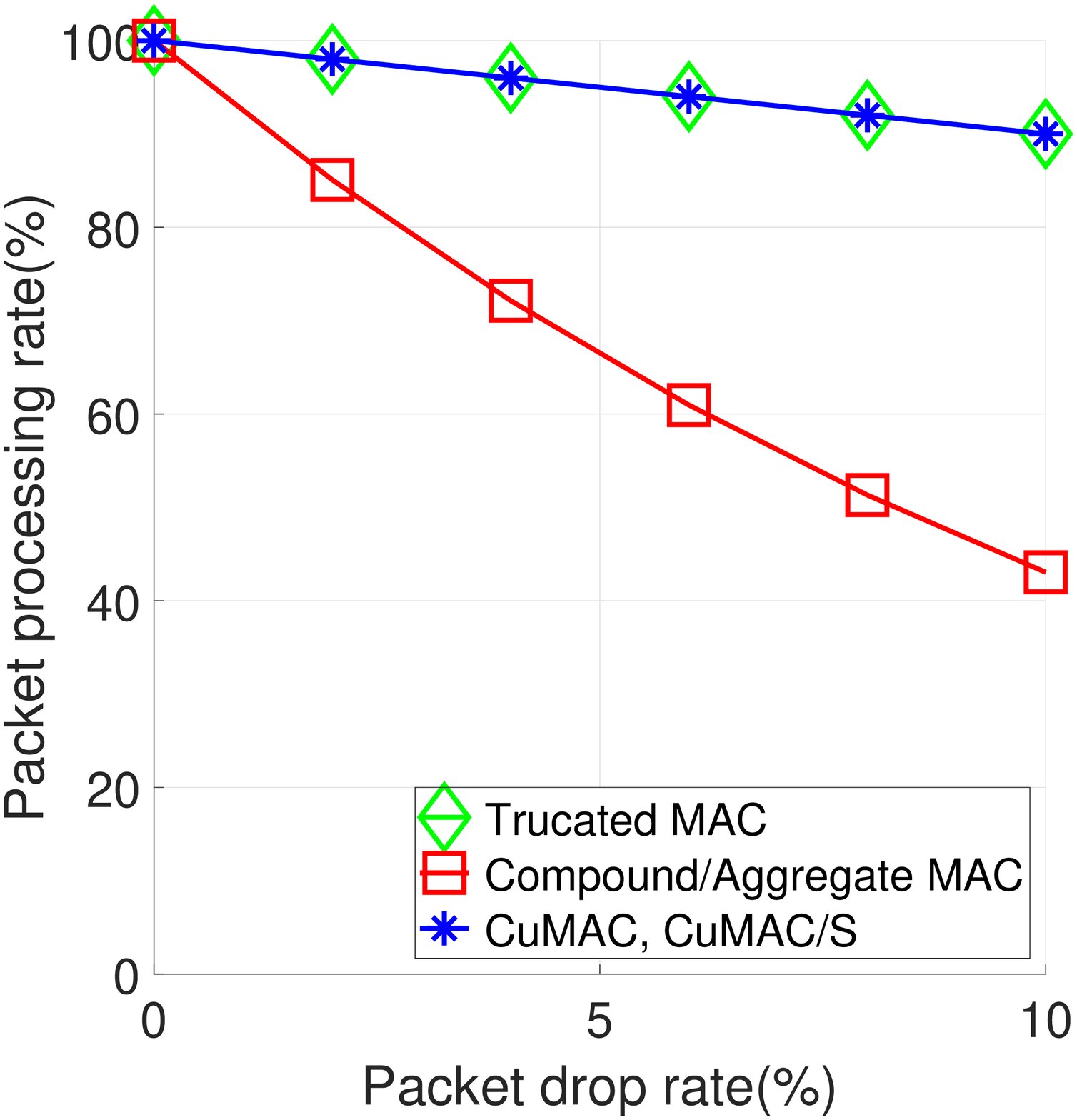}
    \caption{Effect of unreliable communication channel.}
    \label{fig:packetDrop_vs_authentication}
\end{subfigure}
\caption{Illustration of the higher cryptographic strength and higher packet processing rate achieved by CuMAC and CuMAC/S in comparison with the prior art.}
\savesp
\end{figure}

\begin{table*}[t]
\centering
\caption{Comparison of the MAC schemes using the prototype implementation on a real car.}
\label{tab:prototype}
\small
\renewcommand{\arraystretch}{1.2}
\scalebox{0.95}{
\begin{tabular}{|c|c|c|c|c|c|c|c|c|}
\hline
\multirow{2}{*}{\textbf{Scheme}}		& \multirow{2}{*}{\textbf{Code Space}}       		   		& \textbf{Increase in} 				& \multicolumn{2}{c|}{\textbf{Real-Time Auth.}}  & \multicolumn{2}{c|}{\textbf{Full Auth.}} & \multicolumn{2}{c|}{\textbf{Partially Accum. Auth.}} \\ \cline{4-9}
&   & \textbf{Bus Load} 		& \textbf{Delay} & \textbf{Strength} & \textbf{Delay} & \textbf{Strength} & \textbf{Delay} & \textbf{Strength} \\ \hline
Trailing MAC                            & 7410 bytes			    				& 200 \% 			& 3.451 ms  & 0 bit	& 5.616 ms  & 128 bits & 50.000 ms  & 128 bits \\ \hline
Truncated MAC 							& 7410 bytes 							& 8 \%				& 3.440 ms & 16 bits & 3.440 ms  & 16 bits	& 50.000 ms  & 16 bits \\ \hline
Compound/Aggregate MAC  	& 7450 bytes 								& 8 \%				& 3.887 ms & 0 bit & 84.143 ms  & 128 bits & 50.000 ms  & 0 bits	\\ \hline
CuMAC    	    		    & 7522 bytes 							& 8 \% 			& 3.798 ms & 16 bits & 83.983 ms  & 128 bits & 50.000 ms  & 64 bits	\\ \hline
CuMAC/S    	    		    & 7640 bytes 							& 8 \% 			& 3.809 ms & 128 bits & 83.994 ms  & 128 bits & 50.000 ms  & 128 bits	\\
\hline
\end{tabular}}
\end{table*}

\para{Cryptographic Strength.} Figure~\ref{fig:accumulating-security_vs_delay} presents the cryptographic strengths of the MAC schemes versus their authentication delay. In the figure, we observe that CuMAC provides real-time authentication with cryptographic strength of 16 bits, which is the same for the truncated MAC. As more packets are received, partially accumulated authentication is achieved and CuMAC provides increasing cryptographic strength. Finally, CuMAC provides full authentication with cryptographic strength of 128 bits, which is the same as the compound/aggregate MAC. This way, CuMAC enables a receiver to make a trade-off between (accumulated) cryptographic strength and authentication delay. In some latency-tolerant IoT applications, this attribute provides the receiver with operational flexibility to vary the security level and/or packet processing delay based on particular needs of a protocol or rules prescribed by network traffic processing policies. 


Further, in Figure~\ref{fig:accumulating-security_vs_delay}, we observe that CuMAC/S enables the receiver to achieve $128$ bits of cryptographic strength for real-time authentication. In other words, for the messages which can be reliably predicted, the receiver achieves the cryptographic strength of the full authentication without any delay (i.e., immediately after the message is received).

\para{Unreliable Communication Channel.}
The unreliability of the channel is measured by the packet drop rate which is equal to the ratio of the lost packets and the total number of transmitted packets. The performance of each scheme is measured in terms of the packet processing rate which is equal to the ratio of successfully authenticated packets at the receiver and the total number of transmitted packets. 

We evaluate the effect of unreliable communication channel on the MAC schemes in Figure~\ref{fig:packetDrop_vs_authentication}. In the figure, we observe that the packet processing rate in CuMAC and CuMAC/S is equal to that in the truncated MAC. However, the compound/aggregate MAC can enable processing of significantly lower number of packets than CuMAC and CuMAC/S. This is because in compound/aggregate MAC, the verification of a MAC requires the receiver to receive \emph{all} of the packets that contain the messages utilized to compute that particular MAC, and loss of any one of those packets leads to the failure in processing of other packets. For instance, with a typical $10\%$ packet drop rate, the packet processing rate in the compound/aggregate MAC is around $43\%$ which might lead to unacceptable performance in any typical IoT application.


\section{Implementation Results}
\label{sec:implementation}

Here we discuss the results obtained from a prototype implementation of CuMAC and CuMAC/S on a real car.

\subsection{Details of Prototype Implementation}

\begin{figure}[t]
\centering
\includegraphics[width=0.9\linewidth, angle=0]{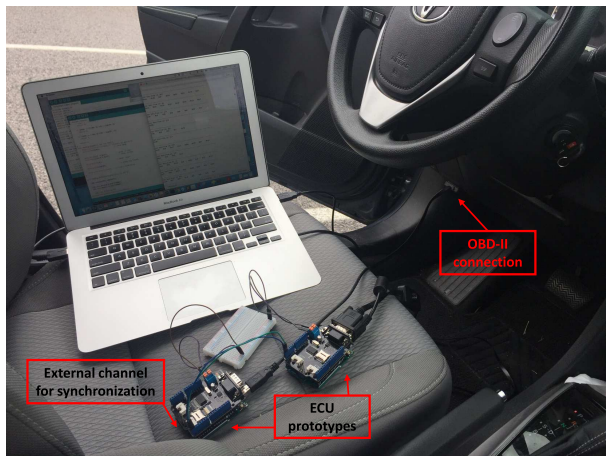}
\caption{Prototype connected to a car's CAN bus.}
\savesp
\label{fig:testbed}
\end{figure}

Figure~\ref{fig:testbed} illustrates the prototype implementation and the setup that were used for running our experiments. The prototype implementation comprised of two ECU prototypes connected to the on-board diagnostics (OBD) port of the CAN bus (with the bus speed of $500$~kbps) of a 2016 Toyota Corolla. The ECU prototype consisted of an Arduino UNO board and a Seeed studio CAN shield. The Arduino UNO board was used to emulate the controller unit of an ECU, and the Seeed Studio CAN shield worked as the interface between the Arduino UNO board and the CAN bus. The Arduino UNO board utilizes an Atmel ATmega328P chip, which includes a low-power 8-bit micro-controller running at 16~MHz clock speed along with a 32~KB flash memory and a 2~KB RAM. These specifications of the ECU prototype are representative of a typical state-of-the-art automotive-grade controller \cite{Mur16}.

With the above experimental setup, we compared six schemes: the trailing MAC, the truncated MAC, the compound MAC, the aggregate MAC, CuMAC, and CuMAC/S. For all schemes, AES-CMAC with a MAC output of 128 bits was utilized as the underlying MAC algorithm. We utilized an open-source cryptography library~\cite{github_aes} to implement AES-CMAC. We found that the average computation time (calculated by averaging the computation time over 1000 executions) of generating a MAC was 0.786 ms. For the truncated MAC, the compound MAC, the aggregate MAC and CuMAC, the size of the tag was set to 16 bits, and the message and tag were inserted into the data field of the same CAN packet. For the trailing MAC, the 128-bit MAC was split into two tags of 64 bits, and inserted into the data fields of two consecutive CAN packets. These packets were transmitted immediately after the CAN packet containing only the message. 

To evaluate the delay performance, we utilized one ECU prototype (called Tx-ECU) to transmit 6-byte messages with the tags on the CAN bus, and another ECU prototype (called Rx-ECU) to measure the end-to-end delay. In the experiment, the Rx-ECU requested the Tx-ECU (through an external synchronization channel) to send a message, and started the timer. The Rx-ECU stopped the timer after verifying the tag and authenticating the message. The delay was measured as the time between starting the timer and stopping the timer. Also, we let the message processing deadline for the message type utilized in the experiment be 50~ms. Note that the processing deadline represents the time within which the authentication tags corresponding to the message are expected to be generated, communicated and verified.

\subsection{Results}

Table~\ref{tab:prototype} summarizes the results from the experiments. The end-to-end delay shown in the table is the worst case delay in processing 1000 CAN messages. The table also presents the cryptographic strengths for real-time, full and partially accumulated authentication in each scheme. From Table~\ref{tab:prototype}, we observe that: (1)~In comparison to other MAC schemes, additional storage is required in CuMAC and CuMAC/S; (2)~Unlike the trailing MAC, CuMAC and CuMAC/S does not increase the bus load significantly; (3)~Unlike the compound MAC and the aggregate MAC, CuMAC and CuMAC/S provide real-time authentication; and (4)~In comparison with the truncated MAC, the compound MAC and the aggregate MAC schemes, CuMAC and CuMAC/S provide significantly higher cryptographic strength for partially accumulated authentication within the processing deadline. 

\section{Conclusion}
\label{sec:conclusion}
We proposed a novel concept for message authentication that we refer to as \emph{cumulative MAC} (CuMAC). CuMAC incurs low communication overhead, and provides high cryptographic strength which is commensurate with the delay in authentication. We also proposed a variant of CuMAC called \emph{CuMAC with speculation} (CuMAC/S) that is more suitable for latency-sensitive applications. Our promising simulation and experimental results validate that CuMAC and CuMAC/S provide significant advantages over the MAC schemes in prior art when deployed in emerging IoT applications, including those that run on energy/bandwidth-constrained networks. 

\appendices

\section{Security Definition}
\label{sec:security_definition}

Here we present the formal security definitions for CuMAC and CuMAC/S. Katz et al. provide the first concrete proof which illustrates that if multiple conventional MACs with cryptographic strength of $\lambda$ bits are aggregated by XOR operation to form an aggregate MAC, then the aggregate MAC is secure with the cryptographic strength of $\lambda$ bits \cite{Eik10,Kat08}. The aggregation procedure employed in CuMAC and CuMAC/S share similar attributes with the scheme proposed by Katz et al. Hence, we present the security definitions which closely follow those presented by Katz et al.

The security evaluation for CuMAC is centered around the notion of unforgeability under chosen message attack with parameter $r$ (uf-cma-$r$), where $r$ indicates the number of packets accumulated for tag verification. We denote by $\mathbf{Adv}_{\mathsf{CuMAC}}^{\textnormal{uf-cma-}r}(\adv,\lambda,q)$, the advantage of the adversary $\adv$ in forging a message for a random key $\mathtt{k}\gets \mathsf{KeyGen}(1^\lambda)$, where $\adv$ can make $q$ queries to the tag generating oracle of CuMAC $O_{\mathsf{CuMAC}}(\mathtt{k},\cdot)$, and verification is performed after accumulating $r$ segments of each MAC. CuMAC is considered to be secure if the advantage of the adversary $\adv$ is negligibly small. Formally, the advantage can be expressed by the probability (represented by $\prob{}$) that the following experiment returns 1.

\smallskip

\noindent $\mathbf{Exp}_{\mathsf{CuMAC}}^{\textnormal{uf-cma-}r}(\adv,\lambda,q)$

\begin{itemize}
\item[] $\mathtt{k}\gets \mathsf{KeyGen}(1^\lambda)$

\item[] Invoke $\adv^{O_{\mathsf{CuMAC}}(\mathtt{k},\cdot)}$ who can make up to $q$ queries to the tagging oracle of CuMAC $O_{\mathsf{CuMAC}}(\mathtt{k},\cdot)$. $\adv$ can query $O_{\mathsf{CuMAC}}(\mathtt{k},\cdot)$ with $n$ arbitrarily chosen messages and receive their CuMAC tags in response.

\item[] $\adv$ outputs a set of $n$ pairs $\left(\set{m_i}_{i=1}^{n},\set{\tau_i}_{i=1}^{n}\right)$.

\item[] Return $1$ if $\mathtt{valid} \gets \mathsf{TagVerify}\left(\mathtt{k},i,m_i,\tau_i \right)$ for all $1\leq i \leq n$, and $\adv$ did not make the query for $m_{i^*}$ to $O_{\mathsf{CuMAC}}(\mathtt{k},\cdot)$, where $i^* = n-r+1$.

\item[] Return $0$ otherwise.
\end{itemize}

\begin{definition}
CuMAC is $(t,q,\epsilon,r)$-uf-cma secure if for any probabilistic polynomial time (PPT) adversary $\adv$ running in time $t$, $\prob{\mathbf{Exp}_{\mathsf{CuMAC}}^{\textnormal{uf-cma-}r}(\adv,\lambda,q)=1}\leq \epsilon$.
\end{definition}

Similar to the experiment $\mathbf{Exp}_{\mathsf{CuMAC}}^{\textnormal{uf-cma-}r}(\adv,\lambda,q)$, the uf-cma-$r$ experiment for CuMAC/S can be readily defined as follows.

\smallskip

\noindent $\mathbf{Exp}_{\mathsf{CuMAC/S}}^{\textnormal{uf-cma-}r}(\adv,\lambda,q)$

\begin{itemize}
\item[] $\mathtt{k}\gets \mathsf{KeyGen}(1^\lambda)$

\item[] Invoke $\adv^{O_{\mathsf{CuMAC/S}}(\mathtt{k},\cdot)}$ who can make up to $q$ queries to the tagging oracle of CuMAC/S $O_{\mathsf{CuMAC/S}}(\mathtt{k},\cdot)$. $\adv$ can query $O_{\mathsf{CuMAC/S}}(\mathtt{k},\cdot)$ with $2n-1$ arbitrarily chosen messages and receive their CuMAC/S tags in response.

\item[] $\adv$ outputs a set of $2n-1$ pairs $\left(\set{m_i}_{i=1}^{2n-1},\set{\tau_i}_{i=1}^{2n-1}\right)$.

\item[] Return 1 if $\mathtt{valid} \gets \mathsf{TagVerify}\left(\mathtt{k},i,m_i,\tau_i \right)$ for all $1\leq i \leq 2n-1$, and $\adv$ did not make the query for $m_{i^*}$ to $O_{\mathsf{CuMAC/S}}(\mathtt{k},\cdot)$, where $i^* = 2n-r$.

\item[] Return 0 otherwise.
\end{itemize}

\begin{definition}
CuMAC/S is $(t,q,\epsilon,r)$-uf-cma secure if for any PPT adversary $\adv$ running in time $t$, $\prob{\mathbf{Exp}_{\mathsf{CuMAC/S}}^{\textnormal{uf-cma-}r}(\adv,\lambda,q)=1}\leq \epsilon$.
\end{definition}

Note that if CuMAC and CuMAC/S are $(t,q,\epsilon,r)$-uf-cma secure for all $r$, then they are also $(t,q,\epsilon)$-uf-cma secure which is the standard notion of security for a MAC scheme. We utilize the aforementioned uf-cma-$r$ security model to define the cryptographic strength for full authentication, partially accumulated authentication and real-time authentication. Note that in the uf-cma-$r$ experiment, when the experiment returns a value of $1$, it implies that $\adv$ is able to forge a valid tag for a packet at which point the receiver has already accumulated $r$ packets. Therefore, if CuMAC or CuMAC/S is $(t,q,\epsilon, n)$ secure, i.e., $r=n$, then CuMAC or CuMAC/S is secure in terms of full authentication. Similarly, if CuMAC or CuMAC/S is $(t,q,\epsilon, r)$ secure for all $2\leq r\leq n-1$, then the scheme is secure for partially accumulated authentication; and if CuMAC or CuMAC/S is $(t,q,\epsilon, 1)$ secure, i.e., $r=1$, then the scheme is secure in terms of real-time authentication. We provide the rigorous proofs of security of CuMAC and CuMAC/S in Appendix \ref{sec:security_proof}. The security of CuMAC and CuMAC/S is based on the following assumption that defines the security of the underlying MAC algorithm \cite{Bel00}.

\begin{assumption}
The underlying deterministic MAC algorithm, $\mathsf{MacGen}$, is $(t,q,\epsilon)$-uf-cma secure---i.e., the probability that an adversary will be successful in producing a forged tag after running for a polynomial time $t$ and making $q$ queries is negligible.
\end{assumption}

\section{Security Proof}
\label{sec:security_proof}

Here we present theorems and corresponding proofs for the security for CuMAC and CuMAC/S. Let CuMAC be instantiated with parameters $(l,n)$, i.e., each MAC is divided into $n$ segments, each of length $l$ bits. Let CuMAC/S be instantiated with parameters $(\beta,l,n)$, i.e., the speculation error rate for the messages is $\beta$, and each MAC is divided into $n$ segments, each of length $l$ bits. Note that in CuMAC and CuMAC/S, the receiver performs real-time authentication by setting $r=1$, partially accumulated authentication by setting $1<r<n$, and full authentication by setting $r=n$.

\begin{theorem}\label{theorem:CuMAC}
For any $t,q \in \NN$ and $\epsilon > 0$, if the underlying deterministic MAC algorithm, $\mathsf{MacGen}$, is $(t,q,\epsilon)$-uf-cma secure, then CuMAC with parameters $(l,n)$ is $(t',q',\epsilon',r)$-uf-cma secure, where 
\begin{align*}
t' \approx t, ~~~~ q' = \frac{q-n+1}{n}, ~~~~ \epsilon' = 2^{l(n-r)} \cdot \epsilon.
\end{align*}

\end{theorem}

\begin{IEEEproof}
Let there be an adversary $\adv$ which succeeds to create a forgery of an authentication tag for CuMAC with a non-negligible probability. We construct a simulator $\sdv$ that interacts with the adversary $\adv$ and creates a forgery of a MAC for the $\mathsf{MacGen}$ algorithm with a non-negligible probability. 

Let CuMAC and the $\mathsf{MacGen}$ algorithm utilize the same secret key $\mathtt{k}$ which is not known to the adversary $\adv$. Also, let the MAC of a message in CuMAC be computed by a query to the tag generating oracle of underlying MAC, which is denoted as $O_{\mathsf{MacGen}}(\mathtt{k},\cdot)$. In this way, $\sdv$ perfectly simulates $O_{\mathsf{CuMAC}}(\mathtt{k},\cdot)$, and hence, the uf-cma-$r$ experiment. Suppose the uf-cma-$r$ experiment for CuMAC returns $1$ with the probability $\epsilon'$ in time $t'$, where an adversary $\adv$ outputs a valid forgery $\left(\set{m_i}_{i=1}^{n}, \set{\tau_i}_{i=1}^{n} \right)$ after $q'$ queries to $O_{\mathsf{CuMAC}}(\mathtt{k},\cdot)$ simulated by $\sdv$. To create a forgery of a MAC for the $\mathsf{MacGen}$ algorithm, the simulator $\mathcal{S}$ proceeds as follows.

For all $i \in [1,n]$ and $i \neq i^*$, the simulator $\mathcal{S}$ queries the $O_{\mathsf{MacGen}}(\mathtt{k},\cdot)$ for the MAC of $m_i$, and obtains the corresponding $\sigma_i$. It divides each MAC into $n$ segments as shown in equation~\eqref{eq:segment}. It recovers the MAC segments of the message $m_{i^*}$ by removing the mask by the MAC segments of other messages as follows:
\begin{align}
\label{eq:sec_rec}
s_{i^*}^k \gets \tau_{i^*+k-1} \oplus \bigoplus_{j=1,j\neq k}^{n}  s_{i^*+k-j}^j.
\end{align}
Since $i^* = n-r+1$, the simulator $\mathcal{S}$ cannot recover the segments $s_{i^*}^k$ with $k \geq r+1$. Hence, it makes a random guess for the rest of the $n-r$ segments, such that $\widetilde{s}_{i^*}^k \sample \set{0,1}^{l}$ for all $k \in [r+1,n]$. Finally, to create the forgery for the underlying MAC algorithm, $\mathsf{MacGen}$, it concatenates all the recovered segments and the guessed segments: $\sigma_{i^*} \gets s_{i^*}^1 || s_{i^*}^2 \cdots || s_{i^*}^r || \widetilde{s}_{i^*}^{r+1} || \cdots \widetilde{s}_{i^*}^{n}$. This means that given a successful forgery of the authentication tag in CuMAC, the probability of creating the forgery of $\mathsf{MacGen}$ is $2^{-l(n-r)}$.

To achieve the forgery of $\mathsf{MacGen}$ as shown above, the simulator $\mathcal{S}$ conducts at most $n \cdot q'$ queries to the $O_{\mathsf{MacGen}}(\mathtt{k},\cdot)$ to reply the $q'$ queries by $\adv$ to $O_{\mathsf{CuMAC}}(\mathtt{k},\cdot)$. Also, the simulator $\mathcal{S}$ conducts $n-1$ queries to $O_{\mathsf{MacGen}}(\mathtt{k},\cdot)$ to obtain $\set{\tau_i}_{i=1,i\neq i^*}^n$. Therefore, if these exists an adversary $\adv$ running in time $t'$ and achieving $\prob{\mathbf{Exp}_{\mathsf{CuMAC}}^{\textnormal{uf-cma-}r}(\adv,\lambda,q')=1}\leq \epsilon'$, then it can be leveraged to create a forgery for the underlying MAC algorithm, $\mathsf{MacGen}$, in time $t'$ plus the time required to evaluate the equation~\eqref{eq:sec_rec}, by making $nq'+n-1$ queries, and with probability $2^{-l(n-r)}\epsilon'$. Hence, if the underlying MAC algorithm, $\mathsf{MacGen}$, is $(t,q,\epsilon)$-uf-cma secure, then CuMAC is $(t',q',\epsilon',r)$-uf-cma secure, where $t' \approx t$, $q' = \frac{q-n+1}{n}$, and $\epsilon' = 2^{l(n-r)}\epsilon$.
\end{IEEEproof}

\begin{theorem}\label{theorem:CuMAC/S}
For any $t,q \in \NN$ and $\epsilon > 0$, if the underlying deterministic MAC algorithm, $\mathsf{MacGen}$, is $(t,q,\epsilon)$-uf-cma secure, then CuMAC/S with parameters $(\beta,l,n)$ is $(t',q',\epsilon',r)$-uf-cma secure, where 
\begin{align*}
t' \approx t, ~~~~ q' = \frac{q-3n+3}{2n-1}, ~~~~\epsilon' = \frac{\epsilon}{(1-\beta)+\beta2^{-l(n-r)}}. \\
\end{align*}
\end{theorem}

\begin{IEEEproof}
Let there be an adversary $\adv$ which succeeds to create a forgery of an authentication tag for CuMAC/S with a non-negligible probability. We construct a simulator $\sdv$ that interacts with the adversary $\adv$ and creates a forgery of a MAC for the $\mathsf{MacGen}$ algorithm with a non-negligible probability. 

Let CuMAC/S and the $\mathsf{MacGen}$ algorithm utilize the same secret key $\mathtt{k}$ which is unknown to the adversary $\adv$. Also, let the MAC of a message in CuMAC/S be computed by a query to $O_{\mathsf{MacGen}}(\mathtt{k},\cdot)$. In this way, $\sdv$ perfectly simulates $O_{\mathsf{CuMAC/S}}(\mathtt{k},\cdot)$, and hence the uf-cma-$r$ experiment for CuMAC/S. Suppose the uf-cma-$r$ experiment for CuMAC/S returns $1$ with the probability $\epsilon'$ in time $t'$, where an adversary $\adv$ outputs a successful forgery $\left(\set{m_i}_{i=1}^{2n-1}, \set{\tau_i}_{i=1}^{2n-1} \right)$ after $q'$ queries to $O_{\mathsf{CuMAC/S}}(\mathtt{k},\cdot)$ simulated by $\sdv$. To create a forgery of a MAC for the $\mathsf{MacGen}$ algorithm, the simulator $\mathcal{S}$ proceeds as follows.

For all $i \in [1,2n-1]$ and $i \neq i^*$, the simulator $\mathcal{S}$ queries $O_{\mathsf{MacGen}}(\mathtt{k},\cdot)$ for the MAC of $m_i$, and obtains the corresponding $\sigma_i$. Additionally, it queries $O_{\mathsf{MacGen}}(\mathtt{k},\cdot)$ for the MAC of the speculated messages $\widehat{m}_i$ and obtains $\widehat{\sigma}_i$ for all $ i \in [i^*+1,i^*+n-1]$ . It divides each MAC into $n$ segments as shown in equation~\eqref{eq:segment}. It recovers the MAC segments of the message $m_{i^*}$ by removing the mask by the MAC segments of other messages as follows:
\begin{align} \label{eq:sec_rec_p}
s_{i^*}^k \gets \tau_{i^*+k-1} \oplus \bigoplus_{j=1,j\neq k}^{n}  s_{i^*-j+k}^j \oplus \bigoplus_{j=2}^{n} \widehat{s}_{i^*+j+k-2}^j.
\end{align}
By following the above procedure, the simulator $\mathcal{S}$ recovers $r$ MAC segments. For all $k\geq r+1$, the simulator $\mathcal{S}$ attempts to recover $s_{i^*}^k$ from the tags received before tag $\tau_{i^*}$ as follows:
\begin{align} \label{eq:sec_rec_p2}
s_{i^*}^k \gets \tau_{i^*-k+1} & \oplus \bigoplus_{j=1}^{n}  s_{i^*-k-j+2}^j  \oplus \bigoplus_{j=2,j\neq k}^{n}  \widehat{s}_{i^*-k+j}^j.
\end{align}
These segments can be recovered with a probability $1-\beta$. If a speculation error occurs, then the corresponding MAC segment is not recovered. In this case, $\sdv$ sets the value of the MAC segment by randomly guessing the bits. Finally, the simulator $\mathcal{S}$ creates a fresh forgery for the underlying deterministic MAC algorithm, $\mathsf{MacGen}$, by concatenating all recovered and guessed segments. The probability that such forgery is correct is $(1-\beta) + \beta\cdot 2^{-l(n-r)}$.

To achieve the forgery of $\mathsf{MacGen}$ as shown above, the simulator $\mathcal{S}$ conducts at most $(2n-1)q'$ queries to $O_{\mathsf{MacGen}}(\mathtt{k},\cdot)$ to answer $q'$ queries by $\adv$ to $O_{\mathsf{CuMAC/S}}(\mathtt{k},\cdot)$. In order to compute operations in equations~\eqref{eq:sec_rec_p} and \eqref{eq:sec_rec_p2}, the simulator $\mathcal{S}$ conducts at most $2n-2$ queries to $O_{\mathsf{MacGen}}(\mathtt{k},\cdot)$ to obtain $\set{\tau_i}_{i=1,i\neq i^*}^{2n-1}$, and at most $n-1$ queries to obtain $\set{\widehat{\sigma}_i}_{i=i^*+1}^{i^*+n-1}$. Therefore, if for an adversary $\adv$ running in time $t'$, we have $\prob{\mathbf{Exp}_{\mathsf{CuMAC/S}}^{\textnormal{uf-cma-}r}(\adv,\lambda,q')=1}\leq \epsilon'$, then we can leverage it to break the underlying MAC algorithm, $\mathsf{MacGen}$, in time $t'$ plus the time required to evaluate equations~\eqref{eq:sec_rec_p} and \eqref{eq:sec_rec_p2}, by making $(2n-1)q'+3n-3$ queries, and with probability $\epsilon' (1-\beta+\beta2^{-l(n-r)})$. Hence, if the underlying MAC algorithm, $\mathsf{MacGen}$, is $(t,q,\epsilon)$-uf-cma secure, then CuMAC/S is $(t',q',\epsilon',r)$-uf-cma secure, where $t' \approx t$, $q' = \frac{q-3n+3}{2n-1}$, and $\epsilon' = \frac{\epsilon}{1-\beta+\beta2^{-l(n-r)}}$.
\end{IEEEproof}

\bibliographystyle{plain}
\bibliography{reference}

\end{document}